\documentclass[letterpaper]{article} 
\usepackage{aaai24}  
\usepackage{times}  
\usepackage{helvet}  
\usepackage{courier}  
\usepackage[hyphens]{url}  
\usepackage{graphicx} 
\usepackage{natbib}  
\usepackage{caption} 
\usepackage{algorithm}
\usepackage{algorithmic}
\usepackage{amsmath}
\usepackage{booktabs}
\usepackage{amssymb}
\usepackage{multirow} 
\usepackage{graphicx} 
\usepackage{threeparttable}
\usepackage{makecell}
\usepackage{array}

\usepackage{mathtools}
\usepackage{bm}
\usepackage{soul}
\makeatletter
\newcommand{\thickhline}{%
    \noalign {\ifnum 0=`}\fi \hrule height 1pt
    \futurelet \reserved@a \@xhline
}

\usepackage{newfloat}
\usepackage{listings}
\DeclareCaptionStyle{ruled}{labelfont=normalfont,labelsep=colon,strut=off} 
\floatstyle{ruled}
\newfloat{listing}{tb}{lst}{}
\floatname{listing}{Listing}
\title{GAMC: An Unsupervised Method for Fake News Detection using Graph Autoencoder with Masking}
\author{
    Shu Yin\textsuperscript{\rm 1},
    Chao Gao\textsuperscript{\rm 1},
    Zhen Wang\textsuperscript{\rm 1*}
}
\affiliations{
    \textsuperscript{\rm 1}Northwestern Polytechnical  University\\

    No. 127, West Youyi Road, Beilin District, \\
    Xi'an City, Shaanxi Province, China
}

\usepackage{bibentry}

\begin{document} 

\maketitle

\begin{abstract}
With the rise of social media, the spread of fake news has become a significant concern, potentially misleading public perceptions and impacting social stability. Although deep learning methods like CNNs, RNNs, and Transformer-based models like BERT have enhanced fake news detection, they primarily focus on content, overlooking social context during news propagation. Graph-based techniques have incorporated this social context but are limited by the need for large labeled datasets. Addressing these challenges, this paper introduces GAMC, an unsupervised fake news detection technique using the Graph Autoencoder with Masking and Contrastive learning. By leveraging both the context and content of news propagation as self-supervised signals, our method negates the requirement for labeled datasets. We augment the original news propagation graph, encode these with a graph encoder, and employ a graph decoder for reconstruction. A unique composite loss function, including reconstruction error and contrast loss, is designed. The method's contributions are: introducing self-supervised learning to fake news detection, proposing a graph autoencoder integrating two distinct losses, and validating our approach's efficacy through real-world dataset experiments.
\end{abstract}

\section{Introduction}
The rapid development of social media has brought immense convenience to people's lives. However, it has also served as a breeding ground for the widespread dissemination of fake news \cite{10.1145/3543507.3583299}. The proliferation of fake news has become a major issue in the digital media era, as it could mislead public perception, affect social stability, and even threaten national political security \cite{10.1145/3395046}. Therefore, the detection of fake news has become a pressing issue that requires swift and effective solutions.

To automatically identify the ever-growing fake news, various approaches have been proposed. Traditional fake news detection methods primarily involve manually designing rules to extract news features and then employing classifiers to categorize these features. However, manually designing rules to extract valuable fake news features could be labor-intensive, and may not always capture the intricacies associated with deceptive content \cite{10.1145/3395046}.

\begin{figure}[htb]
  \centering
  \includegraphics[width=3.3in]{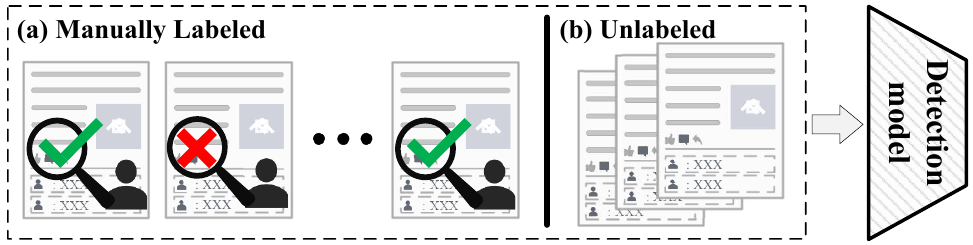}
  \caption{Difference between supervised and unsupervised methods for fake news detection. (a) Existing methods largely rely on manually labeled datasets. The process of manually annotating data is time-consuming, expensive, and often requires expert knowledge to ensure accurate labeling. (b) In contrast, our proposed method, GAMC, is based on unsupervised learning and can directly use unlabeled data for fake news detection. This eliminates the need for costly and time-consuming manual data annotation.}
  \label{fig0}
\end{figure}

In recent years, deep learning has played an increasingly important role in rumor detection, as it can automatically learn and extract underlying patterns and features from large amounts of data, thereby improving the accuracy and efficiency of detection mechanisms \cite{10.1145/3137597.3137600}. For example, Convolutional Neural Networks (CNN) and Recurrent Neural Networks (RNN) have been utilized to learn local and temporal dependencies in text data respectively \cite{li2021survey}. Furthermore, Transformer-based models, such as BERT, have been employed to understand the context and semantic relationships in news articles better \cite{devlin-etal-2019-bert}. These models, pre-trained on large corpora, have shown remarkable success in capturing the complex linguistic characteristics of fake news \cite{9511228}. However, these methods primarily analyze the content of the news, and do not consider the social context information in the process of news propagation \cite{GLAN}.

Recognizing this oversight, researchers have proposed graph-based methods that incorporate social context into the detection process \cite{PSIN}. These methods model the spread of news as a graph, capturing the intricate interactions and relationships among various entities involved in news propagation. However, these supervised methods necessitate large labeled datasets for training as Figure \ref{fig0} \cite{10103675}. The collection and labeling of extensive datasets can be a laborious and resource-intensive task, posing a significant challenge for real-world applications \cite{10.1145/3404835.3463001}.

To address these issues, this paper proposed an unsupervised fake news detection method, GAMC, that employs a Graph Autoencoder with Masking and Contrastive learning. By employing the context and content of the news propagation process as the self-supervised signal, along with a feature reconstruction and contrasting task, this method circumvents the need for labeled datasets.  
Specifically, we first perform data augmentation on the original news propagation graph, which includes random node feature masking and edge dropping, to create two enhanced graphs. Then, we employ the graph encoder to encode these enhanced graphs, yielding the graph-level representation vector. These vectors not only capture the global characteristics of the graph but also contain information about the news propagation process, both its context and content. Once the model is trained, these graph-level representation vectors can be directly used for the task of fake news detection.
Next, we use the graph decoder to map the graph-level representation vector back to the original input, resulting in a reconstructed graph vector. This step is designed to teach the model how to reconstruct the original input from the graph-level representation vector, thus helping the model to better understand and learn the latent patterns of news propagation.
Finally, we design a composite loss function composed of reconstruction error loss and contrast loss. The reconstruction error loss aims to minimize the discrepancy between the reconstructed graph representation and the original graph representation, enabling the graph autoencoder to better learn the latent features of the propagation graph. The contrast loss, on the other hand, ensures that the representations of the two augmented graphs generated from the same propagation graph are as similar as possible after reconstruction.

The contribution of this paper can be summarized as follows:
\begin{itemize}
    \item Self-supervised learning is introduced into the domain of fake news detection, eliminating the dependence on labeled data, which makes the method more applicable to real-world scenarios.
    \item We proposed a graph autoencoder with reconstruction error loss and contrast loss. The reconstruction error loss aims to minimize the discrepancy between the reconstructed and the original graph representations, while the contrast loss ensures that the representations of two augmented graphs, both derived from the same propagation graph, are as similar as possible after reconstruction.
    \item We conducted a series of experiments on real-world datasets, demonstrating the effectiveness of the proposed method. 
\end{itemize}

\section{Related Work}
To provide a comprehensive understanding of the current landscape of our work, we review two primary areas: fake news detection methods and generative self-supervised graph learning techniques.
\subsection{Fake News Detection}
The task of fake news detection can be viewed as a classification problem. The classification process relies on various factors such as the content of the news, the spread pattern, user reactions, and other related data.

Recently, deep learning has been taking an increasingly prominent role in fake news detection. Ma et al. developed a novel recurrent neural network (RNN) based method for rumor detection on microblogging platforms, which outperformed traditional models using hand-crafted features \cite{att-RNN}. Considering the different events, EANN is proposed to effectively extract event-invariant features from multimedia content, thereby enhancing the detection of fake news on newly arrived events \cite{EANN}. To introduce extra knowledge for detecting fake news, Wang et al. proposed a unified framework named KMGCN, using a graph convolutional network to extract textual information, knowledge concepts, and visual information \cite{KMGCN}. However, these methods primarily focus on the content of news, which may fall short in distinguishing ambiguous fake news that is crafted to resemble real news. 

Building on this, researchers begin to explore the potential of leveraging social context information in the process of news propagation. Generally, the social context includes information such as forwarding relationships, comment content, and user preferences, which can provide additional insight into how news spreads in social networks. Bian et al. introduced the bi-directional graph convolutional network that simultaneously captures top-down propagation and bottom-up dispersion features on social media, which enhances traditional deep learning approaches \cite{BiGCN}. To capture rich structural information, GLAN models relationships among source tweets, retweets, and users as a heterogeneous graph,  then effectively encodes both local semantic and global structural information for rumor detection \cite{GLAN}. Considering the influence of user preferences in news propagation, UPFD employs users' historical posts as an endogenous preference, and the news propagation graph as an exogenous context, integrating internal and external information to better identify disinformation \cite{UPFD}. 

However, these supervised methods depend on large labeled datasets. The acquisition of these labeled datasets often requires considerable time, effort, and domain expertise. 

\subsection{Generative self-supervised Graph Learning}
Generative self-supervised graph learning leverages the richly structured data in graphs to learn meaningful representations without the need for explicit labels \cite{wu2021self}. To generate diverse and realistic graphs, Li et al. introduced GraphRNN, a deep autoregressive model that addresses the challenges of graph generation and representation learning \cite{GraphRNN}. Kipf et al. developed a novel Graph Auto-Encoder (GAE) based method that learns to encode a graph into a lower-dimensional space and then decode it back into its original structure, outperforming traditional models using hand-crafted features \cite{VGAE}. Considering that most GAEs lack the ability to reconstruct node features, some work has been dedicated to reconstructing masked features, thereby enhancing the efficiency of self-supervised GAEs in graph representation learning for classification tasks \cite{GraphMAE}. Our work is inspired by the graph mask autoencoder, and we developed a self-supervised graph autoencoder to obtain representations of news for the task of fake news detection.

\section{Problem Definition}
In this paper, the task of fake news detection is to design an automatic discriminator that can learn latent features from a set of unlabeled news. This learned knowledge can then be used to predict the authenticity of unseen news instances. Specifically, the news dataset can be defined as $D=\left\{D_1, D_2, \ldots, D_n\right\}$, where each $D_i$ represents a single news instance in the dataset. Each news instance $D_i$ can be modeled as a graph based on its corresponding propagation process.

Our aim is to learn an unsupervised function, $f$, as defined below:

\begin{equation}
\label{eq1}
f: D \rightarrow Y ,
\end{equation}
where $D$ represents the set of graph representations of the news instances and $Y \in \left\{ F, R\right\}$ (i.e. Fake News or Real News) denotes the set of possible outcomes.

\section{Method}

\begin{figure*}[htb]
  \centering
  \includegraphics[width=1.0\textwidth]{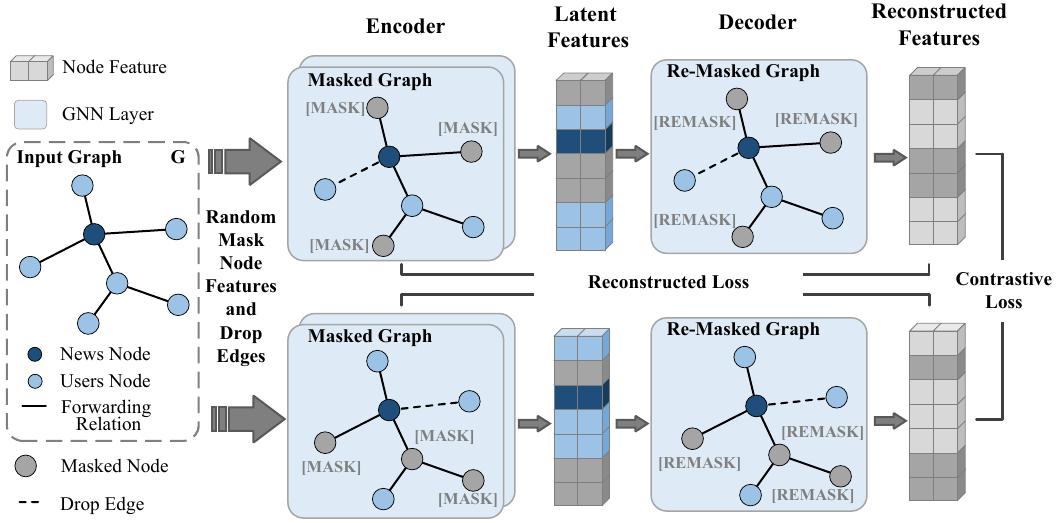}
  \caption{ Overview of the GAMC Fake News Detection Method. The original news propagation graph is first augmented through random feature masking and edge dropping, generating two perturbed graphs. These graphs are then processed through a graph encoder to generate latent representation vectors. A decoder subsequently maps these vectors back to their input space, producing reconstructed feature vectors. Minimizing the composite loss function facilitates the effective learning of the propagation graph's latent features by the autoencoder and ensures the similarity of the reconstructed features across the augmented graphs.
    }
  \label{fig1}
\end{figure*}

In this section, we introduce the GAMC method for fake news detection tasks, designed to capitalize on the inherent context and content of the news propagation process to function as a self-supervised signal, thereby bypassing the need for labeled datasets. As illustrated in Figure \ref{fig1}, the following parts will detail the procedure of employing GAMC for fake news detection, including data augmentation, graph encoding, graph decoding, and the composite loss function.

\subsection{Data Augmentation}
For the GAMC method, data augmentation is an essential first step. It aims to generate augmentation data by transforming the original graph, without altering the underlying semantic context of the news propagation process.

To begin with, each piece of news is modeled into a graph $G = (V, A, X)$, based on the forwarding relationship. $V=\left \{v_n,v_u \right \}$ represents the set of nodes, where $v_n$ is the news node, and $v_u$ signifies the user nodes that forward the news. $A$ represents the adjacency matrix, which embodies the forwarding relationships. $X$ is the feature matrix.  The node feature for the news nodes $v_n$ is the news content embedding, encoded by a pre-trained BERT model, and the node feature for the user nodes $v_u$ is derived from their historical posts, as described in~\cite{UPFD}. The node feature of node $i$ is denoted as $x_{i}$. This graph-based representation can provide a structured view of the news content, making it easier for our model to extract and learn meaningful features. This graph-based representation can provide a structured view of the news content, making it easier for our model to extract and learn meaningful features.

Following the construction of the graph, the data augmentation process employs two strategies: random node feature masking and edge dropping. 

Random node feature masking is the random feature elimination of the nodes when the training begins. Let $V_m \subset V$ be a subset of nodes randomly selected for masking. For each node in $V_m$, we replace its feature vector with a special mask token, denoted as $x_{[MASK]} \in \mathbb{R}^D$. Then, the masked feature matrix $\widehat X$ can be defined as:

\begin{equation}
    \widehat x_{i} =
    \begin{cases}
    x_{[MASK]}, & \text{if } v_i \in V_m \\
    x_{i}, & \text{if } v_i \notin V_m ,
    \end{cases} 
\end{equation}
where $\widehat x_{i}$ is the augmentation feature of node $i$, the augmented
feature matrix $\widehat X$ is constructed by $\widehat x_{i}$.

Edge dropping is the second strategy used for data augmentation. This method disrupts the connectivity of the graph by randomly dropping some edges before training. $E_{Drop}$ is the edge set obtained by randomly sampling from the original edge set $E$, and $A_{Drop}$ denotes the adjacency matrix of $E_{Drop}$. Then the augmented adjacency matrix $\widehat A$ could be calculated as $\widehat A = A - A_{Drop}$. Through the above operations, the augmented graph can be represented as $\widehat G = (V, \widehat A, \widehat X)$. Each of these strategies is applied twice to the original graph, resulting in two distinct augmented graphs, namely $\widehat G_1 = (V, \widehat A_1, \widehat X_1)$ and $\widehat G_2 = (V, \widehat A_2, \widehat X_2)$.

These data augmentation strategies not only ensure the model's effectiveness in the face of complex news propagation patterns but also facilitate the model's learning of the context and content information of the propagation process.

\subsection{Graph Encoding}

Following the data augmentation process, the second integral component of the GAMC method is graph encoding. The purpose of this step is to transform the augmented graphs into a compact and meaningful latent space representation.

The graph encoder in our method is a two-layer Graph Isomorphism Network (GIN)~\cite{GIN}, which is designed to process the augmented graph $\widehat G = (V, \widehat A, \widehat X)$ and generate the graph-level representation vectors. The GIN is selected due to its capacity to capture the topological structure and node features of a graph, making it suitable for the task of fake news detection.

Given an augmented graph $\widehat G$, the GIN encoder operates as follows. At the $l$-th layer, the hidden feature vector $h_i^{(l)}$ for node $i$ is updated using the aggregation function:

\begin{equation}
    h_i^{(l)} = MLP \left( (1 + \epsilon^{(l)}) \cdot h_i^{(l-1)} + \sum_{j\in \mathcal{N}(i)} \left( h_j^{(l-1)} \right) \right),
\end{equation}
where $\mathcal{N}(i)$ is the set of neighboring nodes of $i$, and $h_i^{(0)} = \widehat x_i$ is the input feature vector of node $i$. This process is iteratively conducted for all nodes until the $l$-th layer.

After two layers of information propagation, the GIN encoder outputs a set of node embeddings $H$ for all nodes in the graph. Through the aforementioned encoding step, we can obtain the nodes' latent representations, $H_1$ and $H_2$, for the two augmented graphs $\widehat G_1$ and $\widehat G_2$. These node embeddings will then be pushed into the graph decoder.

While the model has been trained, the graph-level representation vectors obtained from the GIN encoder can also be used directly for news classification tasks. Node embeddings from the encoder are pooled together to generate a graph-level representation vector $F$ for the entire graph:

\begin{equation}
    F = \sum_{i=1}^{n} h_i.
\end{equation}

This graph-level representation vector $F$ captures the overall information of the graph, including both the structural and content information, which are essential for the downstream task of fake news detection.

\subsection{Graph Decoding}
The third core component of the GAMC method is graph decoding. The goal of this step is to map the latent graph-level representation vectors back to the input, namely to obtain reconstructed feature matrices.
Before decoding, we perform a re-mask operation on the masked nodes, forcing the masked nodes to aggregate from their neighbors in order to reconstruct their initial features. For each node in $V_m$, we replace its representation with a special mask token, denoted as $h_{[REMASK]} \in \mathbb{R}^D$. The re-mask representation $\widehat h_{i}$ of $v_i$ can be described as:
\begin{equation}
    \widehat h_{i} =
    \begin{cases}
    h_{[MASK]}, & \text{if } h_i \in V_m \\
    h_{i}, & \text{if } h_i \notin V_m
    \end{cases} .
\end{equation}

The re-masked latent representation $\widehat H$ is constructed by $\widehat h_{i}$. Correspondingly, after the re-mask operation, the hidden representations of graphs $\widehat G_1$ and $\widehat G_2$ can be denoted as $\widehat H_1$ and $\widehat H_2$. Then, we feed the hidden representations of the two graphs into the graph decoder, obtaining the reconstructed features $X_1'$ and $X_2'$.









\subsection{Loss Function}
The loss function in GAMC is to guide the learning process in a way that the difference between the original and reconstructed graphs is minimized, and the contrast between the two reconstructed graphs derived from the same propagation graph is minimized. We define the loss function in two parts: the reconstruction loss and the contrastive loss.

The reconstruction loss aims to ensure the fidelity of the reconstructed feature matrices $X_1'$ and $X_2'$ to the original feature matrices $X_1$ and $X_2$. We use the Mean Squared Error (MSE) between the original and reconstructed feature matrices as reconstruction loss:
\begin{equation}
    \mathcal{L}{rec} = \frac{1}{n} \sum_{i=1}^{n}  \left( | X_1 - X_1' |_2^2 + | X_2 - X_2' |_2^2 \right) ,
\end{equation}
where $n$ is the number of samples.
By minimizing this loss in the training phase, the model could produce reconstructed graphs that closely match the originals, which encourages the graph encoder to learn better graph-level representation vectors.

On the other hand, the contrastive loss is designed to minimize the difference between the two reconstructed graphs derived from the same propagation graph. This is achieved by minimizing the cosine similarity between reconstructed features $X_1'$ and $X_2'$:

\begin{equation}
    \mathcal{L}{con} = \frac{X_1' \cdot X_2'}{\|{X}_1'\| \|{X}_2'\|} .
\end{equation}

By minimizing this loss, the model is encouraged to generate similar representations for one augmented graph.

The overall loss function is then a weighted sum of the reconstruction loss and the contrastive loss:
\begin{equation}
    \mathcal{L} = \mathcal{L}{rec} - \alpha \mathcal{L}{con},
\end{equation}
where $\alpha$ is the hyperparameter that controls the balance between the two loss components.

By minimizing this overall loss, our model is trained to generate robust and discriminative graph-level representations that can be effectively used for the task of fake news detection.

\section{Experiments}
In this section, we validate the effectiveness of the proposed GAMC method by comparing it with some benchmark models on public datasets. Following this, to analyze and validate the necessity of each component in GAMC, we conduct ablation studies. Finally, we investigate the impact of different parameter values within GAMC on the experimental results.
\subsection{Datasets and Settings}
\subsubsection{Datasets}
To validate the efficiency of GAMC, we carried out evaluations on the FakeNewsNet, a published data source for fake news detection~\cite{FakeNewsNet}. This repository is divided into two sub-datasets, PolitiFact and GossipCop. The PolitiFact dataset primarily consists of news related to U.S. politics, while GossipCop is primarily focused on news about Hollywood celebrities. The social context in these two datasets includes the propagation network of news and the history of user comments.  Table~\ref{tab1} provides comprehensive statistics of the PolitiFact and GossipCop datasets.


\begin{table}[htb] 	\footnotesize
	\renewcommand{\arraystretch}{1.0}
	\caption{Statistics of the datasets. In the two datasets, each graph denotes a piece of news.}
	\label{tab1}
	\centering
	\setlength{\tabcolsep}{3.0mm}{
	\begin{tabular}{ccc}
		\toprule	
		\textbf{Dataset} & \textbf{PolitiFact} & \textbf{GossipCop}  \\
		\midrule
            \#News    & 314 & 5464  \\
            \#True News  & 157 & 2732  \\
            \#Fake News  & 157 & 2732  \\
            \#Nodes    & 41054 & 314262  \\
            \#Edges    & 40740 & 308798  \\
            
		\bottomrule
	\end{tabular}}
\end{table}

\subsubsection{Baselines}
We have conducted a comparison of the proposed method GAMC with the following unsupervised methods:

\begin{itemize}
    \item \textbf{TruthFinder} \cite{TruthFinder} is one of the earliest methods for detecting fake news using an unsupervised approach. This method employs an iterative process to determine the veracity of news by assessing the credibility of the source websites from which the news originates.
    \item \textbf{UFD} \cite{UFD} employs a Bayesian network model and an efficient collapsed Gibbs sampling technique. This method leverages users' engagements on social media to understand their opinions regarding news authenticity, capturing the conditional dependencies among the truths of news, users' opinions, and users' credibility.
    \item \textbf{GTUT} \cite{GTUT} is a graph-based method for fake news detection that identifies a seed set of articles, and then progressively labels all articles in the dataset. 
    \item \textbf{UFNDA} \cite{UFNDA} is an unsupervised fake news detection approach. Utilizing a combination of a Bidirectional GRU (Bi-GRU) layer and self-attention within an autoencoder, the method uncovers hidden relationships between features to detect fake news.
    \item \textbf{$($UMD$)^2$} \cite{UMD} is an unsupervised fake news detection framework that encodes multi-modal knowledge into low-dimensional vectors. This method leverages a teacher-student architecture to determine the truthfulness of news by aligning various modalities, then uses them as guiding signals for veracity assessment.
\end{itemize} 

Additionally, we also conducted comparisons with the following classical supervised methods:
\begin{itemize}
    \item \textbf{SAFE} \cite{SAFE} is a multimodal method for fake news detection. It converts images in the news into text, learns the latent representation of text and visual information, then measures the similarity between them to detect fake news.
    \item \textbf{EANN} \cite{EANN} is a multi-modal approach for detecting fake news. It extracts text and image features from news content, and then incorporates an event discriminator using adversarial learning to obtain the event-invariant features of fake news.
    \item \textbf{BiGCN} \cite{BiGCN} learns high-order structural representations from the rumor propagation and dissemination process, while considering the influence of root features, and achieves good performance in rumor detection tasks.
    \item  \textbf{GACL} \cite{GACL} leverages contrastive learning within the loss function to learn the difference between positive and negative samples, and an Adversarial Feature Transform (AFT) module to generate conflicting samples. This approach enhances the model's ability to distinguish event-invariant features, contributing to more robust and efficient fake news detection.
\end{itemize}

\subsubsection{Parameter Settings}
The experiments were conducted on a server equipped with an Intel(R) Xeon(R) Gold 6326 CPU @ 2.90GHz and a GeForce RTX 3090Ti graphics card. The server has 24G of video memory and runs on the Ubuntu 16.04 operating system. We implement the proposed GAMC model using PyTorch. We use accuracy, F1 score, precision, and recall as our evaluation metrics across both datasets. During the data augmentation process, we mask 50\% of the node features and apply dropout to 20\% of the edges. The node features are represented in 768-dimensional space, while the intermediate layer vectors produced by the encoder have a dimensionality of 512. The training procedure consists of 80 epochs, with the Adam optimization algorithm employed to optimize the model. The hyperparameter that controls the balance between the two loss components is set to 0.1. Upon completion of training, we use a Support Vector Machine (SVM) classifier to predict labels, leveraging the graph-level vectors learned by the graph encoder. For the experimental results, we run ten times and take the average values.
\subsection{Overall Performance}
Table \ref{tab2} and Table \ref{tab3} respectively display the performance of the proposed GAMC method and the unsupervised methods. From the results, compared to existing unsupervised methods, GAMC demonstrates noticeable improvements across all four metrics on the two datasets. Specifically, the accuracy on the PolitiFact dataset increased by 3.24\%, and on the Gossipcop dataset, it rose by 18.81\%. This improvement can be attributed to GAMC's unique design, which leverages a graph autoencoder with masking and contrast. The approach harnesses both the context and content of news propagation, thereby providing a more holistic and accurate representation. Additionally, the composite loss function, combining reconstruction error loss and contrast loss, ensures not only that the latent features of the propagation graph are accurately captured but also that the representations of the augmented graphs are closely aligned.

\begin{table}[htbp]
\renewcommand{\arraystretch}{1.25}
\caption{Results of GAMC, compared with unsupervised methods on the PolitiFact dataset. We underline the suboptimal results and bold the top results.}
\centering
\begin{tabular}{c|cccc}
\toprule
\toprule	
 Methods & \textbf{ACC.} & \textbf{Prec.} & \textbf{Rec.} & \textbf{F1.} \\
\midrule
TruthFinder & 0.581 & 0.572 & 0.576 & 0.573 \\
UFNDA & 0.685 & 0.667 & 0.659 & 0.670 \\
UFD & 0.697 & 0.652 & 0.641 & 0.647 \\
GTUT & 0.776 & 0.782 & 0.758 & 0.767 \\
$($UMD$)^2$ & \ul{0.802} & \ul{0.795} & \ul{0.748} & \ul{0.761} \\ \midrule
GAMC & \textbf{0.828} & \textbf{0.825} & \textbf{0.817} & \textbf{0.823} \\
$\textit{variance}$ & $\pm \textit{0.014}$ & $\pm \textit{0.007}$ & $\pm \textit{0.012}$ & $\pm \textit{0.011}$ \\ 
\bottomrule
\bottomrule
\end{tabular}
\label{tab2}
\end{table}

\begin{table}[htbp]
\renewcommand{\arraystretch}{1.25}
\caption{Results of GAMC, compared with unsupervised methods on the GossipCop dataset.}
\centering
\begin{tabular}{c|cccc}
\toprule
\toprule	
 Methods & \textbf{ACC.} & \textbf{Prec.} & \textbf{Rec.} & \textbf{F1.} \\
\midrule
TruthFinder & 0.668 & 0.669 & 0.672 & 0.669 \\
UFNDA & 0.692 & 0.687 & 0.662 & 0.673 \\
UFD & 0.662 & 0.687 & 0.654 & 0.667 \\
GTUT & 0.771 & 0.770 & 0.731 & 0.744 \\
$($UMD$)^2$ & \ul{0.792} & \ul{0.779} & \ul{0.788} & \ul{0.783} \\ \midrule
GAMC & \textbf{0.941} & \textbf{0.935} & \textbf{0.940} & \textbf{0.937} \\
$\textit{variance}$ & $\pm \textit{0.004}$ & $\pm \textit{0.004}$ & $\pm \textit{0.003}$ & $\pm \textit{0.005}$ \\ 
\bottomrule
\bottomrule
\end{tabular}
\label{tab3}
\end{table}

Table \ref{tab4} shows the performance of the proposed GAMC and the supervised methods. Compared with Tables \ref{tab2} and \ref{tab3}, we can observe that supervised methods tend to outperform unsupervised methods. This is primarily attributable to supervised methods taking advantage of the specific label information provided in training datasets, enabling these models to learn more distinctive and discriminative patterns associated with fake news. 

As can be seen in Table \ref{tab4}, on the PolitiFact and Gossipcop datasets, the methods based on news propagation graphs (BiGCN and GACL) perform better than those based on news content (SAFE and EANN). This is due to the rich contextual information embedded within propagation graphs. These methods effectively capture the complex interconnections and behavioral patterns involved in news propagation. In contrast, methods exclusively focused on news content could potentially overlook these significant contextual signals.
Additionally, our unsupervised method GAMC demonstrates superior performance over the classic content-based supervised algorithms, achieving an accuracy improvement of 2.99\% and 12.56\%, respectively. Compared to classic graph-based supervised algorithms, our method shows a minor decrease in accuracy by 4.71\% and 1.06\%. 
However, in real-world scenarios where labeled data may be scarce or costly to obtain, our GAMC method offers an effective alternative. Furthermore, this unsupervised model opens up new possibilities for continual, on-the-fly fake news detection as it can easily adapt to changing data landscapes. As such, the GAMC model not only competes with supervised methods but also provides additional flexibility and cost-effectiveness, making it a robust solution for the challenge of fake news detection.

\begin{table}
\centering
\caption{Results of GAMC, compared with supervised methods on PolitiFact and GossipCop datasets.}
\renewcommand{\arraystretch}{1.25}
\setlength{\tabcolsep}{3mm}{
\begin{tabular}{c|cc|cc}
\hline
\hline
\multirow{2}{*}{Dataset} & \multicolumn{2}{c|}{\textbf{PolitiFact}}             & \multicolumn{2}{c}{\textbf{GossipCop}}                    \\ \cline{2-5} 
                         & \multicolumn{1}{c}{\textbf{Acc.}}   & \multicolumn{1}{c|}{\textbf{F1.}}      & \multicolumn{1}{c}{\textbf{Acc.}}   & \multicolumn{1}{c}{\textbf{F1.}}     \\ \thickhline  
SAFE                  & \multicolumn{1}{c}{0.793}      & \multicolumn{1}{c|}{0.775}         & \multicolumn{1}{c}{0.832}      & \multicolumn{1}{c}{0.811}            \\
EANN                  & \multicolumn{1}{c}{0.804} & \multicolumn{1}{c|}{0.798}  & \multicolumn{1}{c}{0.836} & \multicolumn{1}{c}{0.813}  \\
BiGCN                    & \multicolumn{1}{c}{0.823} & \multicolumn{1}{c|}{0.822}  & \multicolumn{1}{c}{\textbf{0.951}} & \multicolumn{1}{c}{\textbf{0.951}}  \\
GACL                     & \multicolumn{1}{c}{\textbf{0.867}}      & \multicolumn{1}{c|}{\textbf{0.866}}         & \multicolumn{1}{c}{0.907}      & \multicolumn{1}{c}{0.905}       \\
GAMC                     & \multicolumn{1}{c}{\ul{0.828}}      & \multicolumn{1}{c|}{\ul{0.823}}         & \multicolumn{1}{c}{\ul{0.941}}      & \multicolumn{1}{c}{\ul{0.937}}       \\ \hline
\hline
\end{tabular}}
\label{tab4}
\end{table}

\subsection{Ablation Study}

To further elucidate the importance of each component in our proposed GAMC model, we conduct an ablation study in this section. This analysis aims to evaluate the contribution of individual modules by iteratively removing them and observing the effect on the model's performance. We compare GAMC with its various sub-models:
\begin{itemize}
    \item GMAC-Aug removes the data augmentation, including node feature masking and edge dropping.
    \item GAMC-$L_{rec}$ removes the reconstruction loss, and only depends on the contrastive loss to optimize the model.
    \item GAMC-$L_{con}$ removes the contrastive loss while generating only one augmented graph from the propagation graph. 
\end{itemize}
The comparative results of these various sub-models are visually summarized in Table \ref{tab5}.
From the results, it can be observed that:

Removing data augmentation (GAMC-Aug) led to a decrease in accuracy, indicating the importance of this feature in capturing the intricacies of news propagation. Data augmentation in GAMC is instrumental in increasing the autoencoder's feature reconstruction ability. The absence of reconstruction loss (GAMC-$L_{rec}$) made a noticeable difference in performance, weakening the model's ability to accurately regenerate the original graph structure. The contrastive loss helps the model to recognize similarities and differences between different instances, enhancing its discrimination power. By generating only one augmented graph (GAMC-$L_{con}$), the model loses the capability to contrast between various augmented views of the data.
In conclusion, each component of the GAMC model plays a critical role in ensuring optimal performance.

\begin{table}[htb]
\centering
\caption{Results of sub-models of GAMC on PolitiFact and GossipCop datasets.}
\renewcommand{\arraystretch}{1.25}
\setlength{\tabcolsep}{3mm}{
\begin{tabular}{c|cc|cc}
\hline
\hline
\multirow{2}{*}{Dataset} & \multicolumn{2}{c|}{\textbf{PolitiFact}}             & \multicolumn{2}{c}{\textbf{GossipCop}}                    \\ \cline{2-5} 
                         & \multicolumn{1}{c}{\textbf{Acc.}}   & \multicolumn{1}{c|}{\textbf{F1.}}      & \multicolumn{1}{c}{\textbf{Acc.}}   & \multicolumn{1}{c}{\textbf{F1.}}     \\ \thickhline  
GAMC-Aug                 & \multicolumn{1}{c}{0.793} & \multicolumn{1}{c|}{0.791}  & \multicolumn{1}{c}{0.905} & \multicolumn{1}{c}{0.894}  \\
GAMC-$L_{rec}$                    & \multicolumn{1}{c}{0.763} & \multicolumn{1}{c|}{0.758}  & \multicolumn{1}{c}{0.889} & \multicolumn{1}{c}{0.876}  \\
GAMC-$L_{con}$                     & \multicolumn{1}{c}{0.808}      & \multicolumn{1}{c|}{0.793}         & \multicolumn{1}{c}{0.924}      & \multicolumn{1}{c}{0.915}       \\ 
GAMC                     & \multicolumn{1}{c}{0.828}      & \multicolumn{1}{c|}{0.823}         & \multicolumn{1}{c}{{0.941}}      & \multicolumn{1}{c}{0.937}       \\ \hline
\hline
\end{tabular}}
\label{tab5}
\end{table}

\subsection{Parameter Discussion}
To ensure that our proposed GAMC model achieves optimal performance, an investigation and analysis of the parameters was conducted in this section. The mask rate $\lambda$ and edge drop rate $\gamma$ stand out as vital tunable parameters, impacting the model's capacity to understand and process the underlying data structure. For comprehensive insights into their influence, we conduct a series of experiments using both rates varying between 0.1 to 0.9 to encapsulate their entire effective range.
\begin{figure}[ht]
  \centering
  \includegraphics[width=3.3in]{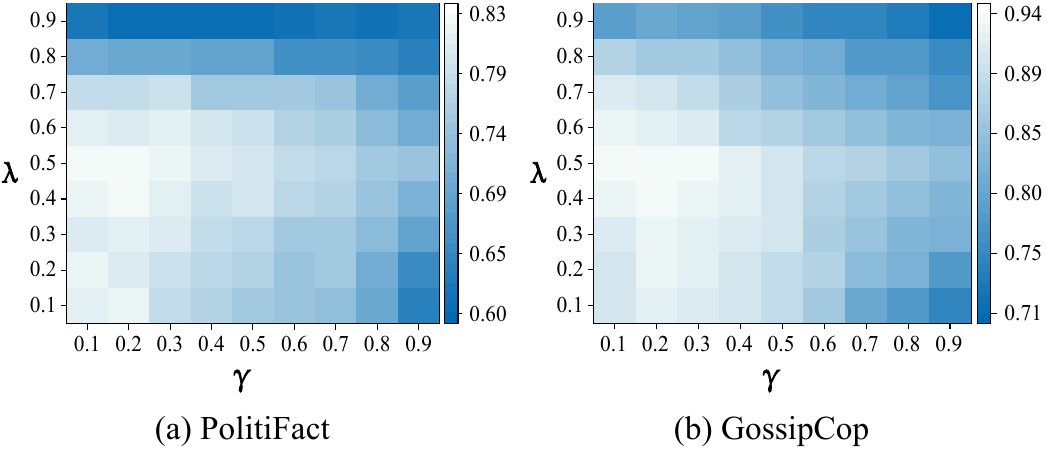}
  \caption{Parameter analysis for $\lambda$ and $\gamma$ in GAMC. (a) shows the results on PolitiFact, and (b) shows the results on Gossipcop.}
  \label{fig2}
\end{figure}
Figure \ref{fig2} shows that when the mask rate is set to 0.5 and the edge drop rate set to 0.2 the result is the best. At a high mask rate, a substantial portion of node information becomes occluded. This leads to the model losing critical information, making it harder to discern patterns and structures essential for its tasks. Samely, a high edge dropping disrupts the inherent structure and connectivity of the original graph. It makes the graph too sparse, thereby losing significant relational data between nodes. On the other hand, the low mask and edge drop rate might not provide enough reconstruction clues. 

\section{Conclusion}
In this study, we introduced GAMC, a novel unsupervised approach to fake news detection. By executing data augmentations like node feature masking and edge dropping, we engender enhanced graphs. Subsequently, we implemented a graph encoding and decoding strategy. Furthermore, the devised composite loss function, including both the reconstruction error loss and the contrast loss, optimally synergizes these components. The reconstruction error loss facilitates the reconstruction of the original graph from its representation vector, thereby strengthening the model's grasp over latent news propagation patterns. The contrast loss facilitates the aligning representations of augmented graphs from the same original graph. Experiments validate that our GAMC method manifests effectively in fake news detection, eliminating the need for extensive labeled datasets, and also beneficial for real-world deployments. In the future, our research will consider multi-modal data sources for even richer representations, and enhance the transparency and interpretability of the model decisions.

\end{document}